# SecGenAI: Enhancing Security of Cloud-based Generative AI Applications within Australian Critical Technologies of National Interest

Christoforus Yoga Haryanto
*School of Science*
*RMIT University*
Melbourne, Australia
s3972281@student.rmit.edu.au

Minh Hieu Vu
*School of Science*
*RMIT University*
Melbourne, Australia
s3985856@student.rmit.edu.au

Trung Duc Nguyen
*School of Science*
*RMIT University*
Melbourne, Australia
s3962594@student.rmit.edu.au

Emily Lomempow
*ZipThought*
Melbourne, Australia
emily@zipthought.com.au

Yulia Nurliana
*ZipThought*
Melbourne, Australia
yulianurliana@zipthought.com.au

Sona Taheri
*School of Science*
*RMIT University*
Melbourne, Australia
sona.taheri@rmit.edu.au

*Abstract*—The rapid advancement of Generative AI (GenAI) technologies offers transformative opportunities within Australia's critical technologies of national interest while introducing unique security challenges. This paper presents SecGenAI, a comprehensive security framework for cloud-based GenAI applications, with a focus on Retrieval-Augmented Generation (RAG) systems. SecGenAI addresses functional, infrastructure, and governance requirements, integrating end-to-end security analysis to generate specifications emphasizing data privacy, secure deployment, and shared responsibility models. Aligned with Australian Privacy Principles, AI Ethics Principles, and guidelines from the Australian Cyber Security Centre and Digital Transformation Agency, SecGenAI mitigates threats such as data leakage, adversarial attacks, and model inversion. The framework's novel approach combines advanced machine learning techniques with robust security measures, ensuring compliance with Australian regulations while enhancing the reliability and trustworthiness of GenAI systems. This research contributes to the field of intelligent systems by providing actionable strategies for secure GenAI implementation in industry, fostering innovation in AI applications, and safeguarding national interests. (*Abstract*)

## I. Introduction

Generative AI (GenAI) technologies offer transformative opportunities for Australia's critical national technologies while introducing unique security challenges [1]. This paper presents SecGenAI, a security framework for cloud-based GenAI applications, focusing on Retrieval-Augmented Generation (RAG) systems. SecGenAI addresses functional, infrastructure, and governance requirements, integrating end-to-end security analysis to generate specifications emphasizing data privacy, secure deployment, and shared responsibility models.

### A. Background on Generative AI

GenAI leverages deep learning on extensive datasets to generate new content [1]. In Australia's Critical Technologies list, AI is recognized as pivotal for innovation [2], [3], [4], [5], [6], but brings security challenges [7], [8], [9], [10], [11], [12], [13], [14]. GenAI systems employ deep learning architectures [15], [16], [17], [18], [19], characterized by high-fidelity outputs, comprehensive data representations, adaptability, and autonomous decision-making.

### B. Importance in the Australian Context

Australian AI guidelines emphasize secure, responsible AI deployment [2], [3], [4], [8]. Public GenAI chatbots face stringent requirements for personal and confidential data [8]. This paper proposes a secure RAG system within the SecGenAI framework to mitigate security risks [20], [21], [22], prioritizing cybersecurity and adhering to principles of confidentiality, integrity, and availability.

### C. The Objective of SecGenAI Framework

SecGenAI aims to develop comprehensive security specifications for cloud-based GenAI applications, focusing on RAG systems, addressing functional, infrastructure, and governance requirements in the context of Australian Critical Technologies [2], [3], [4], [5], [6], [8], [20], [23], [24], [25].

*1) Problem Statements:* This paper addresses five key questions regarding GenAI security, CIA requirements, RAG implementation options, Australian context constraints, and alignment with AI Ethics and Privacy Principles.

*2) Out-of-Scope:* The study excludes public data-reliant GenAI applications, endpoint implementations, highly specialized use cases, non-security technical requirements, and non-AWS cloud platforms.

### D. Definitions

Key terms defined: *1) Functional:* System operations and processes. *2) Infrastructure:* Hardware, software, network, and infrastructural components. *3) Governance:* Policies, procedures, and processes overseeing system operations.

### E. Report Structure

This report is structured as follows: *1) Understanding GenAI Security. 2) Critical Analysis. 3) SecGenAI Framework Requirements Specifications. 4) Discussions and Recommendations.* This structure ensures a thorough exploration of GenAI security challenges and our proposed solutions, building on the work of Lewis et al. [20], Zeng et al. [21], and Zou et al. [22].

## II. UNDERSTANDING GENAI SECURITY

This section covers key concepts, the current state of GenAI security, and RAG's potential to enhance security, highlighting unique challenges like jailbreaking attacks and data leakage.

### A. Foundational Concepts in Generative AI Security

GenAI uses massive datasets to generate outputs from user prompts [1]. RAG integrates information retrieval to enhance contextual relevance [20]. GenAI often uses GANs [15], [16] and Transformer models [17]. Foundation Models (FMs) serve as a basis for specialized applications [26]. Security in these systems focuses on protecting data from unauthorized access, disclosure, and disruption [27], emphasizing confidentiality, integrity, and availability [28], [29].

In Australia, Critical Technologies of National Interest impact national interests, defense, economy, and social cohesion [4], [5], [6]. AI development is guided by Australian Privacy Principles [30], AI Ethics Principles, and AI Ethics Framework [2], [3], [8].

AI's evolving nature introduces new security challenges [24], including dual-use potential [23]. Specific threats include jailbreaking attacks [14], data leakage [9], adversarial attacks [15], [16], and model poisoning [14]. Countermeasures include homomorphic encryption [31], Zero Trust Architecture [32], data masking [33], and differential privacy techniques [34], [35]. Additional strategies involve continuous authentication [36] and Attribute-Based Access Control (ABAC) [37].

Cloud environments add complexities and opportunities for enhanced security and scalability [38]. Major providers offer secure AI deployment solutions, implementing the Shared Responsibility Model [39], [40], [41].

### B. The Current State of GenAI Security

To understand GenAI security, we must compare it with traditional security [14], summarized below:

TABLE I. NEW CHALLENGES IN GENAI

| Challenge | Explanation | Comparison to Traditional Security |
|---|---|---|
| Jailbreaking Attacks | Work around the prohibited content generation rules. | Traditional systems face root access breaches while GenAI face uses manipulation [9]. |
| Prompt Injection Attacks | Malicious inputs produce unintended outputs. | SQL injections manipulate databases; prompt injections manipulate AI responses [14]. |
| Data Leakage Risks | GenAI may reveal sensitive data unintentionally. | Traditional data leaks occur through breaches; GenAI can leak without a breach, such as using an inversion attack [7]. |
| Generation of Insecure Code | Generate code with vulnerabilities due to imperfect understanding. | Traditional tools do not autonomously generate code; AI introduces new vectors for flaws in the codebase [10]. |
| Use by Threat Actors | Enhances cyber-attacks, like crafting sophisticated phishing emails. | Traditional tools lack AI's automation and adaptability, making GenAI a potent tool for scalable attacks [11]. |

Yao et al. [13] and Zhu et al. [14] analysed where traditional countermeasures fall short:

TABLE II. SHORTCOMINGS OF TRADITIONAL COUNTERMEASURES TO CHALLENGES IN GENAI

| Aspect | Traditional Countermeasures | Shortcomings |
|---|---|---|
| Emergent Threat Vectors | Relies on known issues, security analysis, and enumerating system capabilities. | GenAI may leverage undocumented issues and unexpected capabilities in the existing system on a large scale [12]. |
| Expanded Attack Surfaces | Secures known data inputs and interactions, focusing on well-defined perimeters. | User-generated datasets for training and inference, including private ones, can be manipulated [14]. |
| Deep Integrations | Uses isolation techniques like sandboxing to compartmentalise and secure components. | GenAI's interconnected nature makes traditional isolation ineffective [13], [14]. |
| Economic Values | Focuses on protecting systems from common, economically motivated attacks. | GenAI is usually deployed on higher economic value functionality, further exacerbating the risk [14]. |
| Sophistication of Attacks | Uses rule-based system, signature, and ML classifier to detect attack | GenAI can be used to craft highly sophisticated attacks that may already consider existing rules and bypass the detection of the existing tools [12]. |

This analysis highlights key issues such as jailbreaking attacks [14], prompt injection attacks, heightened data leakage risks, insecure code generation, and malicious use of GenAI tools. Traditional security measures are inadequate for addressing the expanded attack surfaces, deep integration challenges, and economic incentives for targeting GenAI systems [12], [13], [14]. Additionally, GenAI enhances and executes attacks on both traditional and other GenAI systems. Threat actors can use GenAI for crafting convincing phishing emails, automating malicious code generation, and performing sophisticated data poisoning attacks [9], [14].

### C. Enhancing Security through RAG

RAG models offer a promising solution to these seven GenAI security challenges: *1) Robust Information Retrieval:* Uses vetted, reliable sources, reducing risks from unmoderated internet sources [20], [42]. *2) Dynamic Response Capabilities:* Can update databases and models to counter emerging threats [12], [42]. *3) Controlled Data Flow:* Allows close monitoring from retrieval to generation, enhancing security checks [13], [42]. *4) Protection from Data Poisoning:* Curated sources mitigate poisoning risks common in models trained on unfiltered datasets [14]. *5) Modular Architecture:* Independently secures retrieval and generative components, isolating potential breaches [43]. *6) Data Governance and Compliance:* Supports stringent governance by controlling data sources, aiding compliance with regulations like Australia's Privacy Act [44]. *7) Auditability and Transparency:* Enhances transparency and auditability through separation of retrieval and generation, meeting AI governance requirements [45]. Public LLMs often lack these controls, being trained on vast public data, which may include sensitive or biased information [46], [47]. Though RAG is not without security concerns (discussed in section III.A), its architecture provides a strong foundation for addressing many GenAI security challenges.

### D. Related Works in GenAI Security

This section reviews key publications and frameworks in GenAI security, highlighting their relevance and limitations for addressing the unique challenges of RAG systems and their security:

TABLE III. SUMMARY AND COMPARISON OF VARIOUS WORKS IN GENAI SECURITY

| Publication Name | Key Focus | Limitations in RAG Security |
|---|---|---|
| OWASP LLM AI Cybersecurity and Governance Checklist [45] | Comprehensive guide for LLM application security | Lacks RAG-specific guidance and implementation details |
| CSIRO AI Ethics Principles in Practice: Perspectives of Designers and Developers [44] | Implementation of Australian AI Ethics Principles | Does not address RAG-specific challenges and trade-offs |
| A Study on the Implementation of Generative AI Services Using an Enterprise Data-Based LLM Application Architecture [48] | Framework for RAG-based GenAI services | Insufficient coverage of crucial security aspects (e.g., data privacy, monitoring, compliance) |

The reviewed publications [44], [45], [48] underscore the need for RAG-specific security measures and highlight the necessity for an "end-to-end" security analysis covering functional, infrastructure, and governance aspects. While these frameworks provide valuable insights into AI security and ethics, they often lack detailed guidance on implementing and securing RAG systems, pointing to the critical need for tailored strategies and further research.

## III. CRITICAL ANALYSIS

This chapter examines the security aspects of GenAI applications, focusing on RAG systems in the context of Australian Critical Technologies of National Interest. Our analysis reveals eight key security challenges: *1) Vulnerability to Novel Attacks:* RAG models are vulnerable to embedding inversion, attribute inference, and membership inference attacks [7], jailbreaking attacks [9], [49], and prompt injection attacks [22], [49]. *2) Expanded Attack Surface:* Massive untrusted datasets increase data poisoning risks [22]. *3) Deep Integration Risks:* Unmediated integration with powerful systems creates attractive targets [19], while traditional access control and isolation are challenging to apply. *4) Data Leakage:* Complex models may inadvertently leak sensitive data [19], [42]. *5) Malicious Use:* Strong generative capabilities can be abused to create harmful content at scale [14], [19]. *6) Challenges in Applying Traditional Security Measures:* Monolithic architecture complicates the application of standard security practices [19]. *7) Incomplete Solutions for Privacy and Bias:* RAG reduces but does not eliminate these issues [42]. *8) Persistent Hallucinations:* RAG may still generate problematic content absent from source documents [42], [50].

These challenges have four significant implications for Australian Critical Technologies of National Interest: *1) APP Compliance:* Organizations must actively protect personal information in RAG systems, especially to fulfil APP 11 [30]. Emphasis is needed on preventing private data leakage from both knowledge bases and training data [21], [30], [49], [51]. *2) Framework Alignment:* Strong coordination between functional, infrastructure, and governance aspects is crucial [2], [3], [7], [8], [16], [22], [25], [42], [52], [53]. *3) Dual Use and Misuse Potential:* RAG models could generate harmful content, necessitating oversight [4], [5], [6], [11], [12], [23]. *4) Ethical Considerations:* Bias, fairness, transparency, and accountability must be addressed at model and system levels [2], [5], [6], [26], [44], [47].

This analysis forms of the SecGenAI framework, which addresses these challenges comprehensively.

### A. Functional Security Analysis

Our functional security analysis is based on the seminal work by Lewis et al. [20] titled "Retrieval-Augmented Generation for Knowledge-Intensive NLP Tasks". This paper has become a foundational reference for various RAG implementations by leading technology companies [54], [55], [56], [57], [58], [59]. We also examined two other RAG models [42], [50]. Our methodology is inspired by Schneider et al. [60] and augmented by GPT-4's capabilities in static analysis [61].

The mapping of each component to security concerns based on the design of RAG by Lewis et al. [20] is as follows: *1) Query Input:* Susceptible to injection attacks if not properly handled [14]. *2) Query Encoder:* Risks include unauthorized access, data poisoning, or tampering [22], [49]. *3) Document Retrieval:* Can expose sensitive information without robust encryption and access controls [19]. *4) Maximum Inner Product Search:* Sensitive vectors could lead to data inference attacks if not securely managed [19]. *5) Seq2Seq Model:* Vulnerable to model inversion attacks and manipulation [7]. Generative parts are also vulnerable to hallucinations [42], [50]. *6) Marginalisation Process:* Data aggregation can expose patterns or sensitive information if not handled securely [7], [21]. Further, the above components and concerns are mapped to the root cause, listed in the table:

TABLE IV. ANALYSIS OF COMPONENTS, SECURITY CONCERNS AND ROOT CAUSES IN RAG SYSTEMS

| Component | Description of Process | Security Concerns | Root Cause |
|---|---|---|---|
| Query Input | The user inputs a query into the system. | Sensitive information leakage, and injection attacks. | Improperly handled input data, lack of robust validation mechanisms. |
| Query Encoder | Encodes the user query. | Unauthorized access, and data tampering. | Inappropriate encryption, and poor access controls. |
| Document Retrieval | Retrieves documents by the encoded query. | Data leakage, unauthorized access. | Inappropriate encryption, and weak access controls. |
| Maximum Inner Product | Identifies relevant documents. | Inference attacks, sensitive information exposure. | Weak data protection measures, and lack of data masking techniques. |
| Seq2Seq Model | Generates output from documents. | Model inversion attacks, manipulation of model. | Insufficient model training, and lack of adversarial training. |
| Marginaliza-tion Process | Combines outputs into final prediction. | Data aggregation attacks, information leakage. | Poor data aggregation methods, and lack of secure processing. |

Based on the above root causes, we can categorize them and propose solutions as follows:

*1) Identity and access management:*

   *a) Query Input:* Validate and sanitize inputs to secure against injection attacks.

b) *Document Retrieval:* Implement strong authentication and authorization controls for sensitive information to prevent unauthorized access.

  2) *Private data confidentiality and integrity:*

  a) *Query Encoder:* Use encryption and secure data handling protocols to protect encoded data from interception or misuse.

  b) *Maximum Inner Product Search:* Implement rigorous data integrity safeguards to prevent data inference from sensitive vectors.

  3) *Generative model security:*

  a) *Seq2Seq Model:* Detect and prevent model tampering and ensure robust model design to resist adversarial attacks.

  b) *Marginalisation Process*: Secure data aggregation to prevent exposure of data patterns or sensitive information, and protect against inference attacks and leakage.

  These three categories address the root causes identified and will form the basis for generating the SecGenAI functional requirement specifications in Chapter IV. The functional security analysis reveals critical vulnerabilities in the RAG system components, emphasizing the need for robust security measures. These findings have direct implications for the infrastructure setup detailed in the next section.

B. *Infrastructure Security Analysis*

  This section explores the infrastructure security of RAG systems based on Intel's fast RAG infrastructure model [56] consisting of: *1) Private Knowledge Base:* Stores raw data [56]. *2) Vector Database:* Repository for processed data [56]. *3) Input/Output Guard Railing:* Manages prompts and returns answers, detecting harmful prompts and preventing prompt injections. [56]. *4) Embedding Model and Retrieval Vector Search:* Transform raw data into refined vectors stored in the vector database [56].

  The primary goal of designing an IT infrastructure for GenAI and RAG is to satisfy the CIA triad: *1) Confidentiality:* Keep data private and accessible only by authorized entities. Components with sensitive data should be in separate network environments to prevent unauthorized access [29]. *2) Integrity:* Ensure data remains accurate and unaltered during processing and storage. Use encryption techniques like AES for data at rest and TLS for data in transit [27]. *3) Availability:* Ensure data and systems are available to authorized users when needed. Implement redundancy and failover mechanisms to maintain service continuity [63]. To achieve these goals, components should be on different servers with minimal access rights to others.

  To ensure data confidentiality: *1) At-Rest Data Encryption:* Encrypt the vector database and private knowledge base using AES [27]. *2) In-Flight Data Encryption:* Use RSA or SSL/TLS with HMAC for data movement to ensure a secure channel [64].

  To ensure data integrity: *1) Hashing and Digital Signature:* Verify data integrity during transit and storage using hashing techniques and digital signatures [65]. *2) Appropriate Placement of Integrity Measure:* Apply digital signatures to data exchanges between systems and users.

  To ensure availability [63]: *1) Denial-of-Service Attack Mitigation:* Use firewalls to detect and drop malicious connections. *2) Disaster Recovery Plans:* Establish protocols to restore services quickly after an attack. *3) Automated Recovery:* Continuous monitoring with logs for analysis and automated recovery of overloaded or shutdown servers.

  With a cloud system design, achieving the CIA triad is more efficient. Amazon Web Services (AWS) offers comprehensive solutions, while other providers like Azure or Google Cloud Platform offer equivalent functionality. This paper uses AWS as an example.

  Infrastructure Methods for RAG Security consist of: *1) Sandbox GenAI Infrastructure:* Divide infrastructures into clusters across availability zones. Use access control to create a private network connecting servers. *2) Database Infrastructure Connection:* Control access for GenAI and RAG components connecting to databases and enable encryptions. *3) Network Security Settings:* Separate public and private network zones. *4) External Attack Prevention:* Detect and filter network attacks using suitable services and automate accordingly. *5) Disaster Recovery and Incident Response:* Implement data and instance backup and recovery and automate accordingly. Detailed infrastructure requirement specifications will be outlined in Chapter IV.B.

C. *Governance Framework Analysis*

  1) *ISO 38500 Evaluate-Direct-Monitor (EDM):* We employ this approach to ensure reliability and bias-free sources [66] even if the sources are not academic in nature.

  a) *Evaluate:* Assess current governance frameworks, identify gaps, and analyze their applicability to GenAI.

  b) *Direct:* Provide directives on implementing governance frameworks effectively.

  c) *Monitor:* Assess current governance frameworks, identify gaps, and analyze their applicability to GenAI.

  2) *Evaluating the Australian Guidelines*

  The strengths of current Australian guidelines include *1) A Comprehensive Risk Management Framework:* ACSC's recommendations emphasize systematic threat assessment and mitigation [25], *2) Holistic Ethical Integration:* Australia's AI Ethics Principles incorporate security with other ethical dimensions [2], *3) Public Sector Focus:* Clear directives for data privacy and protection [8], and *4) Shared Responsibility Model:* Clear segregation of responsibilities between vendors and users.

  Whereas, its limitations include: *1) Lack of Technical Specificity:* ACSC guidance often lacks detailed technical specifications [25], *2) Rapidly Evolving Threat Landscape:* AI threats evolve quickly, necessitating frequent updates, *3) Broad and Inclusive Principles:* Can be difficult to translate into concrete security measures [2]. *4) Interim Nature:* Lacks a long-term strategy and doesn't fully address risks associated with generative AI [8].

  3) *Evaluating Casey-Alvarenga's SRM*

  The Shared Responsibility Model (SRM) divides responsibilities between Cloud Service Providers (CSPs) and customers with some shared responsibilities [67], [68]. Noting some source reliability and bias: *1) Fundamental of SRM*: Provide foundational knowledge but may lack practical insights [67], [68]. *2) Cloud Service Provider SRM:* Offer detailed models but may emphasize strengths over weaknesses [39], [40], [41]. *3) Third-Party Expert and Consulting:* Provide balanced views but may have their own biases [69], [70], [71], [72], [73]. Shared responsibilities include security configurations, compliance, encryption, and

IAM collaboration [69]. Responsibilities vary by service type (IaaS, PaaS, SaaS) [73]. The following table outlines the comparison of SRM across CSPs:

TABLE V. THE SIMILARITIES AND DIFFERENCES OF SRM ACROSS CSPs

|  | AWS | Microsoft Azure | Google Cloud |
|---|---|---|---|
| Similarities | Providers secure the cloud infrastructure and customers secure their specific usages of the cloud. | | |
| Differences | Focuses on "security of the cloud" vs. "security in the cloud" [39]. | Introduces "shared fate," emphasizing collaborative risk management [40]. | Integrates AI-specific security layers: AI platform, application, and usage [41]. |

SRM offers several strengths, including clear division of responsibilities, out-of-the-box compliance, and defined incident response protocols [72]. However, it also has limitations such as overlapping responsibilities [70], lack of transparency from CSPs, and challenges with complex workloads [71].

## IV. SecGenAI Framework Requirements Specifications

These specifications build on top of the critical analysis and propose the minimum requirements for securely implementing GenAI applications within Australian Critical Technologies of National Interest, assuming all basic cloud-based application requirements are already met.

### A. Functional Requirements Specifications

Building on Chapter III's analysis of GenAI security challenges, this section outlines specific functional requirements to mitigate these risks. The requirements are categorized under Identity and Access Management, Data Confidentiality and Integrity, and Model Security.

TABLE VI. IDENTITY AND ACCESS MANAGEMENT functional requirements.

| Functional Requirements | Description | Rationale |
|---|---|---|
| Continuous Authentication | The system shall implement continuous authentication mechanisms to monitor and validate user identities based on behavioural patterns throughout each session. | GenAI enables a new threat vector that includes deep fakes, which are very hard to detect. Continuous authentication makes it harder for deep fakes to establish their identity consistently [74], [75]. |
| Adaptive Authentication | The system shall consider the usage context and behaviour associated with ongoing sessions when providing access to the data source for the RAG. | While currently not yet fully standardised, adaptive authentication can use the context provided in the interaction itself to adapt the access management [76]. |
| Attribute-Based Access Control (ABAC) | The system shall use fine-grained access control to ensure access between the user, the application, and the data store. | Provides finer control over access permissions, allowing for dynamic and context-sensitive security measures [77]. |

TABLE VII. PRIVATE DATA SECURITY AND INTEGRITY functional requirements.

| Functional Requirements | Description | Rationale |
|---|---|---|
| Homomorphic Encryption | The system shall keep the data encrypted using homomorphic encryption while doing secure computations on the data. | The vector representation of the documents used as the data source for RAG can be kept in an encrypted format [31], [78], [79], [80]. |
| Data Masking and Tokenisation | The system shall implement data masking and tokenisation techniques to protect sensitive information within the data source. | Data masking and tokenization help obfuscate sensitive information while preserving the usability of the data for legitimate processing [81]. |
| Data Integrity Verification | The system shall implement mechanisms to verify the integrity of the data source regularly. | Regular data integrity verification ensures that the data has not been tampered with or corrupted, maintaining its accuracy and reliability for organisational use. Ordinary techniques such as hashes are employed for this purpose along with artificial fingerprinting to detect the source of the data [82], [83]. |

TABLE VIII. GENERATIVE MODEL SECURITY functional requirements

| Functional Requirements | Description | Rationale |
|---|---|---|
| Adversarial Attack Mitigation | The system shall implement techniques to detect and defend against adversarial input designed to deceive the machine learning model. | Adversarial attacks can manipulate model outputs, leading to incorrect and harmful responses [15], [52], [84]. |
| Model Parameter Encryption | The system shall encrypt model parameters to preserve model security. | Encryption protects model parameters from being exposed or tampered with to reduce the risk of model theft and unauthorized modifications [78], [80]. |
| Secure Model Training | The system shall use secure protocols for training and updating machine learning models, ensuring the confidentiality and integrity of training data and model parameters. | Secure training environments protect the model from data poisoning attacks and ensure the integrity of the training process. Techniques such as differential privacy and federated learning enhance security [35]. |

### B. Infrastructure Requirements Specifications

Building on the infrastructure security analysis in Chapter III.B, this section outlines key aspects of infrastructure requirements.

#### 1) Sandbox GenAI Infrastructure

Deploy GenAI applications in a sandboxed environment using containerization tools like Docker or virtualization platforms [26]. With cloud infrastructure, set up different instances across various nodes for isolation.

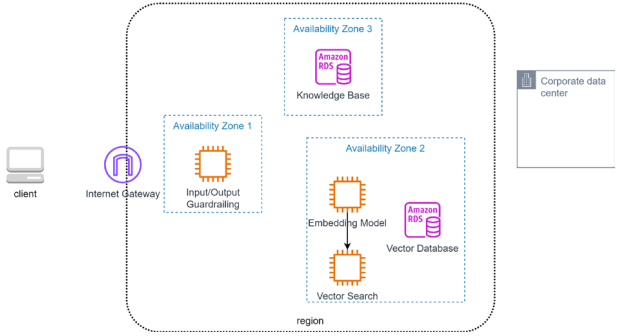

Fig. 1. High-level topology of RAG infrastructure

To achieve sandboxing for GenAI application infrastructure, it should be placed in a dedicated Availability Zone (AZ). Additionally, the connection of GenAI infrastructure with other organizations can be limited using several measures: *1) Design a Virtual Private Network:* Segregate the network into smaller subnets to separate outbound from inbound infrastructure. This isolation helps control damage during network disruptions, allowing operators to easily identify and recover failed subnets. *2) AWS Security Groups:* control data flow by allowing specific network ports and protocols. This prevents attackers from using outbound infrastructure to inject malicious scripts into inbound databases [85]. *3) Applying Identity and Access Management:* Restrict access to authorized entities only, adding an extra layer of security and ensuring proper connections to each infrastructure.

*2) Database Infrastructure Connection*

Use a read replica, asynchronously replicated from the main database in a different Availability Zone.

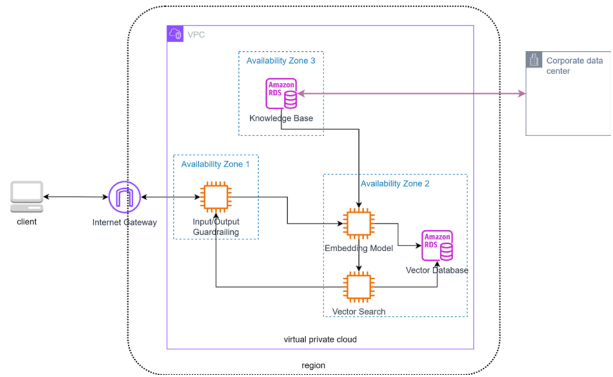

Fig. 2. High-level topology of RAG database connection

Implement strict authentication and authorization protocols using AWS Identity and Access Management (IAM). For data encryption, AWS offers two services: *1) AWS Key Management Services (KMS):* Serverless managed service for cryptographic key management [86]. *2) AWS CloudHSM:* Offers cryptographic key provisioning via dedicated hardware security modules [87].

*3) Network Security Settings*

Network connections need to be segregated: *1) External Connections:* Manage user access with Security Groups, controlling allowed ports and protocols [88] *2) Internal Connections:* Segment internal hosts and services into different subnets within the same VPC.

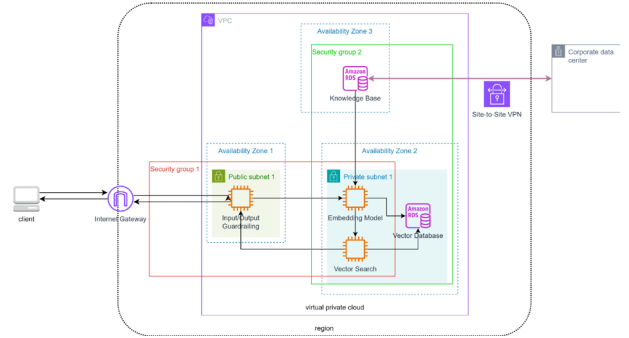

Fig. 3. Network topology for RAG infrastructure

*4) External Attack Prevention*

Use AWS Firewall Manager to define rules for DDoS attacks and log activities [89]. Add a Web Application Firewall (WAF) and Firewall Manager to filter traffic, block, and monitor malicious activities with streamlined administration [16], [90], [91]. AWS Kinesis Data Firehose can streamline data generated by Firewall Manager, storing it in AWS S3, further processing it using AWS Glue and Athena, and then eventually displaying it in Grafana [92], [93], [94], [95].

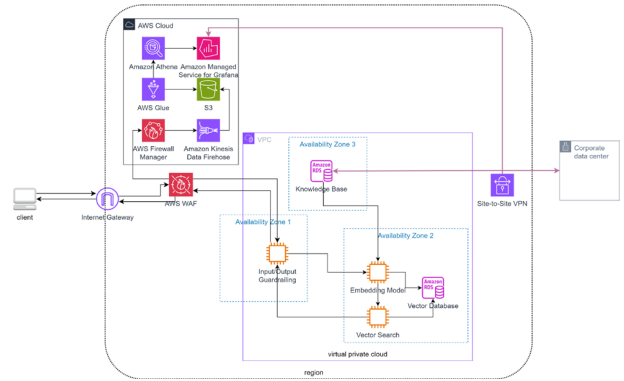

Fig. 4. Network topology with attack prevention

*5) Data Backup and Disaster Recovery*

In AWS, EC2 and RDS instance storage can be snapshotted and stored in AWS S3 [96], [97]. Recovery Time Objective (RTO) and Recovery Point Objective (RPO) are key metrics for disaster recovery plans [98]. Consider a warm standby disaster recovery plan for less critical systems or a multi-site active plan for better RTO and RPO. Regular security exercises and simulations are vital to improve incident response and disaster recovery efficiency [99].

*C. Governance Requirements Specifications*

*1) AI Governance Principles and Requirements:* based on ISO 38500 EDM Evaluation cycle [66]: *a) Ensure Fairness:* Create inclusive AI systems adhering to anti-discrimination laws [2], [25], [100], [101]. *b) Ensure Accountability:* Hold all parties accountable throughout the AI lifecycle with human oversight [2], [102]. *c) Ensure Content Origin and Traceability:* Use digital watermarking and cryptographic provenance [101], [102], [103]. *d) Ensure Data Protection and Privacy:* Prioritize safeguarding user data, aligning with APP, GDPR, and CCPA [2], [30], [102]. *e) Regular Security Audits and Penetration Testing:* Conduct ongoing audits to identify vulnerabilities [103]. *f) Ensure Reliability and Safety:* Continuously monitor AI systems for accuracy, reproducibility, and safety [2], [100], [101], [102]. *g) User Support and Consent Management:* Provide training and clear consent mechanisms [103]. *h) Third-Party Risk Management:* Evaluate and manage third-party security practices. *i) Ensure Transparency and Explainability:* Provide clear, context-specific explanations [2], [100], [102], [103]. *j) Compliance and Legal Responsibility:* Comply with relevant laws and sector-specific regulations. *k) Continuous Monitoring, Improvement, Community Engagement and Feedback:* Regularly update AI systems and engage with users [66], [103].

*2) AI Shared Responsibility Model*

Distinguishes responsibilities between an organization and its AI service provider [104]. Supports IaaS, PaaS, and SaaS users with varying levels of security management.

*3) Shared Responsibility Model for Cloud-Based Generative AI*

The obligations of AI providers as well as the responsibilities of users: *a) Cloud Service Provider (CSP) Responsibilities:* Infrastructure security, network security, data security, platform security, service security, monitoring and logging tools, *and* compliance with industry standards, *b) Customer (Organization) Responsibilities:* Data classification, access controls, encryption key management, application security, configuration management, network security, monitoring and incident response, compliance with regulations, *c) Shared Responsibilities:* Identity and access management, security of operating systems and applications, data encryption, effective communication channels, security policy implementation, continuous monitoring.

*4) Governance Implementation Directive*

Following the ISO 38500 EDM cycle [66]: *a) Define Roles and Responsibilities:* Document security responsibilities for both CSP and customer [70], [71], *b) Understanding and awareness:* Ensure comprehension of security obligations [72], *c) Establish Communication Channels:* Set up effective protocols for security issues, *d) Implement Security Policies:* Develop and review security policies addressing various aspects, *e) Continuous Monitoring and Improvement:* Regularly review and update security practices, conduct assessments and audits. This closes the loop of the entire ISO 38500 EDM cycle [66] as part of the Monitoring activity.

*5) Evaluation of the Governance Requirements*

The shared responsibility model for cloud-based GenAI services outlines CSP and customer roles, offering: *a) Comprehensive Governance:* Covers infrastructure security, data security, platform security, compliance, incident response, data management, application security, network security, monitoring, and governance, *b) Clarity of Roles:* Defines unique and joint responsibilities, fostering a cooperative approach and clear accountability, *c) Addressing the Fundamentals:* Encompasses fairness, accountability, content origin, data protection, reliability, transparency, compliance, and continuous improvement, *d) Iterative Improvements:* Emphasizes ongoing monitoring, community involvement, and adaptability to evolving security risks.

Challenges in implementation include complexity, keeping up with emerging threats, and compliance with changing regulations. Overall, this model provides a strong framework for secure and responsible GenAI implementation.

V. DISCUSSIONS AND RECOMMENDATIONS

This discussion evaluates how the SecGenAI framework mitigates critical security challenges in GenAI systems through functional, infrastructure, and governance measures.

*A. SecGenAI Framework as the End-to-End Solution*

The SecGenAI Framework addresses security gaps in cloud-based GenAI, particularly RAG technology, by delivering: *1) Functional Requirements:* Ensuring identity and access management, data integrity through homomorphic encryption, data masking, and model security safeguards. *2) Infrastructure Requirements:* Securing cloud infrastructure for RAG, isolating components, securing connections, preventing attacks, and ensuring business continuity with AWS services. *3) Governance Requirements:* Aligning GenAI with ethical principles and regulatory standards, clearly defining roles and obligations of CSPs and customers.

*B. SecGenAI Framework Implementation Spectrum*

Implementation can vary based on the organization's risk profile: *1) Full Model:* "Ideal security" for large enterprises and high-risk organizations. *2) SaaS Model:* "Multi-tenant" for small-to-medium enterprises. *3) Hybrid Model:* Combines elements of both, balancing security, cost, and efficiency. Each approach has trade-offs in terms of security, cost, and complexity.

*C. Future Works*

While this report focuses on GenAI security, future work should address cost-effective multi-tenant solutions for small businesses and the development of a prioritization matrix for hybrid solutions.

TABLE IX. COMPARISON TABLE BETWEEN IDEAL SECURITY DESIGN AND MULTI-TENANT DESIGN

| Aspects | Design A: "Ideal security" | Design B: "Multi-tenant" |
|---|---|---|
| Description | The scheme is suitable for large enterprises and follows all the best practices in the industry to ensure a strict and high level of security and compliance. | This design follows the practice of "as-a-service", in which the user organisations will not take responsibility for the infrastructure from the provider. |
| Security level | High and compliant. Also, it is reasonably compliant with APP 11. | Medium and reasonably compliant. Not trivial to comply with APP 11. |
| Implementation cost | High. | Small for users, higher for providers initially |
| Operational cost | High. | Metered for users, fixed for providers. |
| Use cases | Organizations with sensitive data, large enterprises, and government entities. | Small organizations, market research organisations, and small retailers. |

*D. Recommendations*

Based on the analysis, we recommend the following '4A' approach::

1) *Adopt:* Australian organisations should adopt the SecGenAI Framework as a comprehensive guide for securing cloud-based GenAI systems.

2) *Adapt:* Choose between "ideal security design," "multitenant design," or a hybrid approach based on specific needs and constraints.

3) *Adept:* Build proficiency in secure GenAI implementation through continuous monitoring, regular audits, and updates.

4) *Advance:* Encourage collaboration and information sharing among organisations, industry bodies, and government agencies to refine and adapt the SecGenAI framework over time.

## VI. CONCLUSION

In conclusion, the SecGenAI framework provides comprehensive and actionable recommendations for enhancing the security of cloud-based GenAI systems, particularly those using RAG technology. By adopting this framework and following the evidence-based recommendations, organizations can effectively mitigate risks, ensure compliance, and realize the transformative potential of GenAI securely and responsibly. This paper serves as a valuable resource not just for implementation, but also for guiding policymakers, and engaging stakeholders in the development of secure GenAI systems. However, it is important to recognize that GenAI security is an ongoing process requiring continuous vigilance, adaptation, and collaboration. Therefore, further research and development will be essential to refine and extend the SecGenAI framework as GenAI and the threat landscape continue to evolve.

## VII. ACKNOWLEDGEMENTS

This work was part of a class project report made possible through RMIT University's Industrial Awareness Project in collaboration with our industry partner. We would like to thank Dr. Amy Corman and Dr. Mahshid Sadeghpour for their guidance and support throughout this project.